\documentclass[conference]{IEEEtran}
\IEEEoverridecommandlockouts
\usepackage{cite}
\usepackage{amsmath,amssymb,amsfonts}
\usepackage{algorithmic}
\usepackage{graphicx}
\usepackage[export]{adjustbox}
\usepackage{textcomp}
\usepackage{xcolor}
\usepackage{url}
\usepackage{listings}
\usepackage{hyperref}

\def\BibTeX{{\rm B\kern-.05em{\sc i\kern-.025em b}\kern-.08em
    T\kern-.1667em\lower.7ex\hbox{E}\kern-.125emX}}
\begin{document}

\title{Optimizing Branch Predictor for Graph Applications\\
}

\author{
\IEEEauthorblockN{Upasna}
\IEEEauthorblockA{\textit{Dept. of Computer Science and Engineering} \\
\textit{Indian Institute of Technology Ropar}\\
Rupnagar, India \\
upasna.21csz0002@iitrpr.ac.in}
\and
\IEEEauthorblockN{Venkata Kalyan Tavva}
\IEEEauthorblockA{\textit{Dept. of Computer Science and Engineering} \\
\textit{Indian Institute of Technology Ropar}\\
Rupnagar, India \\
kalyantv@iitrpr.ac.in}

}

\maketitle

\begin{abstract}
Real-world graph applications are generally larger than the size of the cache itself. Due to this reason, the memory hierarchy was identified as a key bottleneck by the earlier works. Undoubtedly, the performance can be achieved by improving cache, there is still a scope for performance gain by improving branch prediction accuracy. In graph processing applications, the occurrence of branch mispredictions is very frequent and is a major limitation for the overall performance. Within a program, there are different kinds of branches that recur throughout its execution. Although lots of branch predictors (BP) have been developed earlier to capture the static and dynamic behavior of branches. Branch predictors can yet be further optimized to handle the branches that cause mispredictions. 

\end{abstract}


\section{Introduction}
The efficiency of modern microprocessors highly depends upon the parallelism exhibited by the programs. But, it is not always possible due to control dependencies, data dependencies, and name dependencies. \cite{r1}. To improve the instruction-level parallelism, one of the important components of modern microprocessors is the branch predictor. It detects branches and predicts the branch outcome. Branch Target Buffer (BTB) gives the branch target address in the early pipeline stages. By predicting the branch outcome, the number of wasted clock cycles due to control hazards can be reduced. Correct predictions can lead to performance improvement. Mispredictions by the branch predictor will lead to both performance drop and increased energy consumption because of the pipeline flush and correct-path restoration. Hence it is highly desirable for the branch predictor to have high accuracy~\cite{r3}. 


In graph applications, the conditional statements are heavily dependent on graph characteristics like the degree of graph, number of nodes, number of neighbouring nodes, number of in/out edges, weight of edges, etc.,. Every node in the real world graphs can have different values for these parameters making it difficult for the branch predictor to learn the correct pattern from simply observing the history. Further, for conditional statements like {\tt for} loop, the iteration count may vary widely depending on these graph characteristics. Regular graph like a binary tree, still has some pattern~\cite{b0} but every graph need not be regular and their topology varies drastically based on the data they represent. Due to these reasons significantly high branch misprediction rates are observed in case of graph applications~\cite{b0}. Accordingly, in this work, we analyze the behavior of branches of the popular graph applications. We first identify a subset of branches that contribute to the maximum fraction of the total branch occurrences. We notice that among these high occurring branches, few are biased and remaining unbiased. Prediction for biased branches is trivial, from their history. For the unbiased branches though, prediction is hard because of the above mentioned characteristics of the graphs.


Branches are inherent to any program. Processing of large sized graph applications involves lots of branches. Branch mispredictions lead to significant power and performance loss. The unbiased conditional branches having random branch outcomes make it difficult for the branch predictor to learn a pattern and provide correct prediction. In the branches like sentinel controlled loops where number of iterations are not known well in advance, learning the branch outcome pattern is difficult. This provides a scope to further improve the prediction accuracy of branch predictors while processing these graph applications.    

Branch prediction is not a new problem and there are numerous predictors  proposed so far in the literature. We find that there are limited studies on branch predictors specifically considering graph applications. A detailed study of the performance of branch prediction for graph processing applications was done by Ahmed et al. \cite{b0}. In their comparative analysis considering branch predictors, like, One-bit, Pentium-M \cite{b11}, TAGE (TAgged GEometric length predictor) \cite{b16} and Perfect branch predictor (theoretical, having 100\% accuracy), one of the major claims is that no matter how simple or how sophisticated branch predictor, the branch predictor performance is almost similar in the case of graph applications. 

For our work, we observe the performance of various branch predictors, such as One-bit Predictor, Loop, Local, Global, Neural Network based Branch Predictor, Pentium-M  \cite{b11}, TAGE \cite{b16} \cite{b20} \cite{b13}, Perceptron \cite{b15}, and Piecewise Linear Branch Predictor (PLBP) \cite{b21} \cite{b22} for graph workloads. Section IV presents the graphs showing IPC and MPKI comparison with these branch predictors. We find that Piecewise Linear Branch Predictor (PLBP) performs well but it still has scope of improvement as compared to perfect branch predictor. So, we use sophisticated hashing mechanism to prevent aliasing and use features such as current PC and last N-PCs (N=4, found empirically) for indexing into weight table of PLBP. We further observe the impact of these optimizations on sorted graph datasets.     


The rest of the paper is organized as follows. Section II describes the simulator and benchmarks used for the experiments. Section III shows critical branches in graph workloads. Section IV presents our work and describes our proposal. Finally, Section V concludes the paper.  \\

\section{Experimental Setup}
In order to analyze the control instructions of graph applications and to  study the branch prediction techniques, we use Sniper Simulator \cite{b1} and GAP (Graph Algorithm Platform) Benchmark Suite \cite{b2}. The baseline configuration of Sniper simulator is shown in Table~\ref{fig36}.  
\begin{table}[t]
  \centering
  \caption{Baseline Configuration}
  \label{fig36}
  \begin{tabular}
  {|l|p{6cm}|}
    
    \hline Core & 16 cores, core model = nehalem, type = interval, frequency = 2.66GHz, dispatch width = 4 \\
    \hline L1-I/D Cache & 32 KB size, 64B block size, L1-I 4-way set associative and L1-D 8-way, data access time = 4 cycles \\
    \hline L2 Cache & 256 KB size, 64B block size, 8-way set associative, data access time = 8 cycles  \\
    \hline L3 Cache & 8 MB size, 64B block size, 16-way set associative, data access time = 30 cycles \\
    \hline DRAM & access latency = 45 cycles, queuing model \\
    \hline 
  \end{tabular}
\end{table}

\subsection{GAP Benchmark Suite}

GAPBS contains Breadth-First Search ({\tt BFS}), PageRank ({\tt PR}), Connected Components ({\tt CC}), Betweenness Centrality ({\tt BC}) and Triangle Counting ({\tt TC}) graph kernels. These kernels are used very commonly and are representative of many real world applications within social networks, science and engineering, data mining, etc. The set of kernels in GAP benchmark suite is computationally diverse as it includes both traversal-centric and compute-centric kernels \cite{b2}. We omit Single-Source Shortest Paths ({\tt SSSP}) because of its inconsistent results. 

\subsubsection*{Extending the GAPBS}

The in-built datasets in GAPBS (like {\tt road}, {\tt twitter}, etc.,) are prohibitively large for simulation, so we extend the GAP benchmark with smaller datasets, namely, {\tt webGoogle (web)}, {\tt amazon (am)}, {\tt roadCA (ro)}, {\tt wiki-talk (wi)} and {\tt cite-patents (ci)}  \cite{b9}. The characteristics of these datasets, namely, number of vertices, number of edges, average degree are provided in Table~\ref{fig16}. 


\begin{table}[t]
  \centering
  \caption{Graph Datasets included in GAPBS}
  \label{fig16}
  \begin{tabular}
  {|l|c|c|p{1cm}|p{2.5cm}|}
    \hline \textbf{Dataset} & \textbf{Vertices} & \textbf{Edges} & \textbf{Avg. Degree} & \textbf{Description} \\
    \hline amazon & 410.2K & 3.4M & 8 & E-Commerce\\
    \hline roadCA & 1.9M & 5.5M & 2 & Road network of California \\ 
    \hline webGoogle & 916.4K & 5.1M & 5 & Web graph from Google \\
    \hline wiki-talk & 2.3M & 5M & 2 & Wikipedia Network\\
    \hline cite-patents & 6M & 16M & 2 & Citations made by patents \\
    \hline \end{tabular}
\end{table}


\section{Critical branches in Graph Workloads}

\begin{figure}[h]
	\vspace{-2mm}
	\centering
	\includegraphics[keepaspectratio, height=4.5cm, frame] 
        {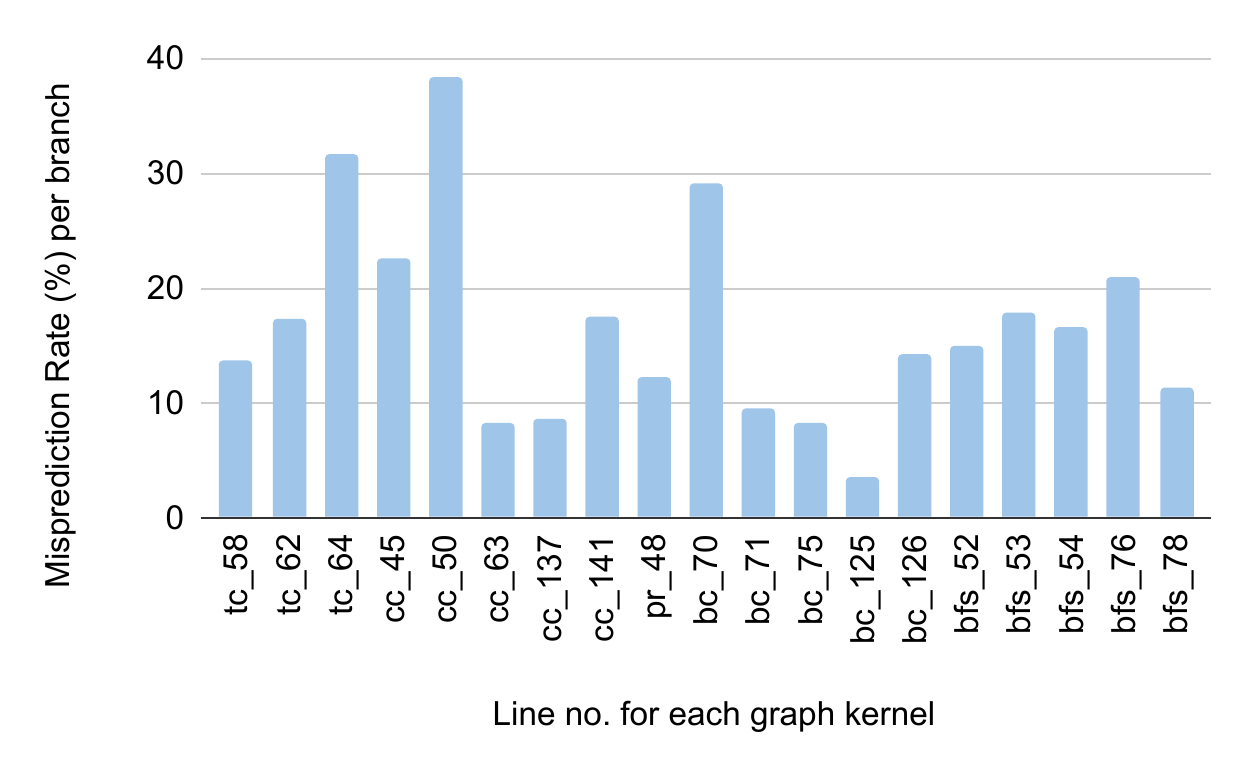}
	\vspace{-2mm}	
	\caption{Miss Rate of critical branches for all graph kernels executed with amazon dataset. [x-axis shows kernel\_lineNo] 
    }
	\label{fig10}
	\vspace{1mm}
\end{figure}

To identify the branches contributing to high misprediction rate, we study GAPBS workloads and collect per branch stats such as the total occurrence of each branch, number of times the branch is taken and not taken, the actual outcome of the branch for a given set of kernel and dataset combination, and use these stats to identify the high frequency branches that contribute maximum to the branch miss rate. We call such branches as the critical branches. We find these critical branches as unbiased and showing high misprediction rate. We identify a branch as biased if number of times a branch is taken (or not taken) divided by the total occurrences of the branch is less than 10\%, otherwise we classify it as an unbiased branch. Figure~\ref{fig10} shows the miss rate of these individual critical branches of all the graph kernels executed with {\tt am} dataset. We calculate the misprediction rate of an individual branch as percentage of total number of the branch mispredictions divided by total number of occurrences of the branch. The x-axis in Figure~\ref{fig10} shows kernel\_lineNo, where the line numbers are of critical branches in GAPBS graph kernels. 


1. TC (Triangle Counting)
\begin{lstlisting}[language=C++]
56   for (NodeID u=0; u < g.num_nodes(); u++) {
57    for (NodeID v : g.out_neigh(u)) {
58*       if (v > u)
...
61       for (NodeID w : g.out_neigh(v)) {
62*         if (w > v)
...
64*         while (*it < w)
\end{lstlisting}

2. CC (Connected Components)
\begin{lstlisting}[language=C++]
43   NodeID p1 = comp[u];
44   NodeID p2 = comp[v];
45*   while (p1 != p2) {
...
48     NodeID p_high = comp[high];
...
50*     if ((p_high == low) ||
         (p_high == high && compare_and_swap(comp[high], high, low)))
...
62   for (NodeID n = 0; n < g.num_nodes(); n++) {
63*     while (comp[n] != comp[comp[n]]) {
...
137*       for (NodeID v : g.out_neigh(u, neighbor_rounds)) {
...
141*       for (NodeID v : g.in_neigh(u)) {
\end{lstlisting}

3. PR (Page Rank)
\begin{lstlisting}[language=C++]
46     for (NodeID u=0; u < g.num_nodes(); u++) {
47       ScoreT incoming_total = 0;
48*       for (NodeID v : g.in_neigh(u))
49         incoming_total += outgoing_contrib[v];
\end{lstlisting}

4. BC (Betweenness Centrality)
\begin{lstlisting}[language=C++]
70*         for (NodeID &v : g.out_neigh(u)) {
71*           if ((depths[v] == -1) &&
        (compare_and_swap(depths[v], static_cast<NodeID>(-1), depth))) {
...
75*           if (depths[v] == depth) {
...
125*         for (NodeID &v : g.out_neigh(u)) {
126*           if (succ.get_bit(&v - g_out_start)) {
\end{lstlisting}


5. BFS (Breadth First Search)
\begin{lstlisting}[language=C++]
51   for (NodeID u=0; u < g.num_nodes(); u++) {
52*     if (parent[u] < 0) {
53*       for (NodeID v : g.in_neigh(u)) {
54*         if (front.get_bit(v)) {
...
76*       for (NodeID v : g.out_neigh(u)) {
77         NodeID curr_val = parent[v];
78*         if (curr_val < 0) {
\end{lstlisting}


The code snippets provided in Listings 1-5 having line numbers marked with $*$ are the critical branches having high misprediction rate. Most critical branches depend upon the no. of vertices, in and out neighbours, but all such branches are not critical. The critical branch is unbiased and has high number of occurrences. We find that these critical branches contribute to the maximum fraction ($>$98\%) of the total branches. These critical branches play a vital role in branch predictor performance shown in graphs in rest of the paper.     


\section{Our Work}
\begin{figure}[h]
	\vspace{-2mm}
	\centering
	\includegraphics[width=0.4\textwidth, height=4cm, frame]{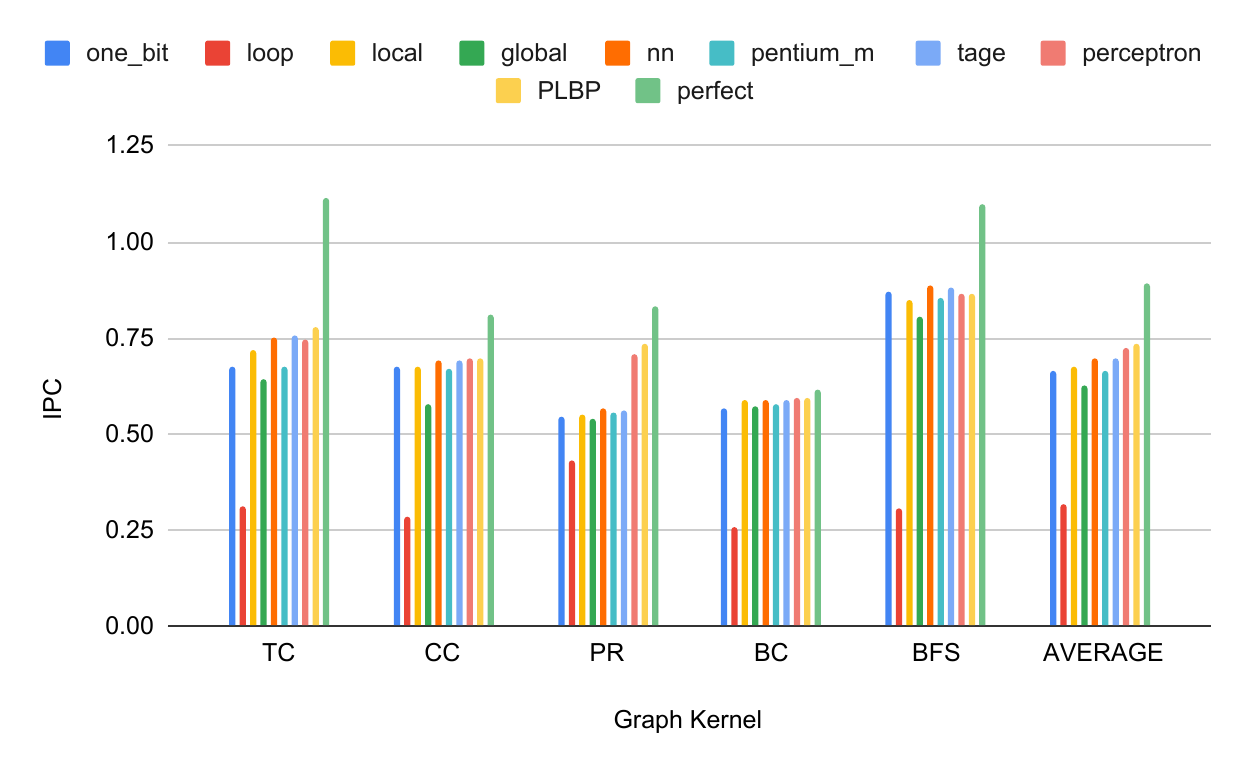}
	\vspace{-2mm}	
	\caption{IPC observed when using various Branch Predictors for all graph kernels average over the three datasets amazon, roadCA, webGoogle}
	\label{fig11}
	\vspace{1mm}
\end{figure}

\begin{figure}[h]
	\vspace{-2mm}
	\centering
	\includegraphics[width=0.4\textwidth, height=4cm, frame]{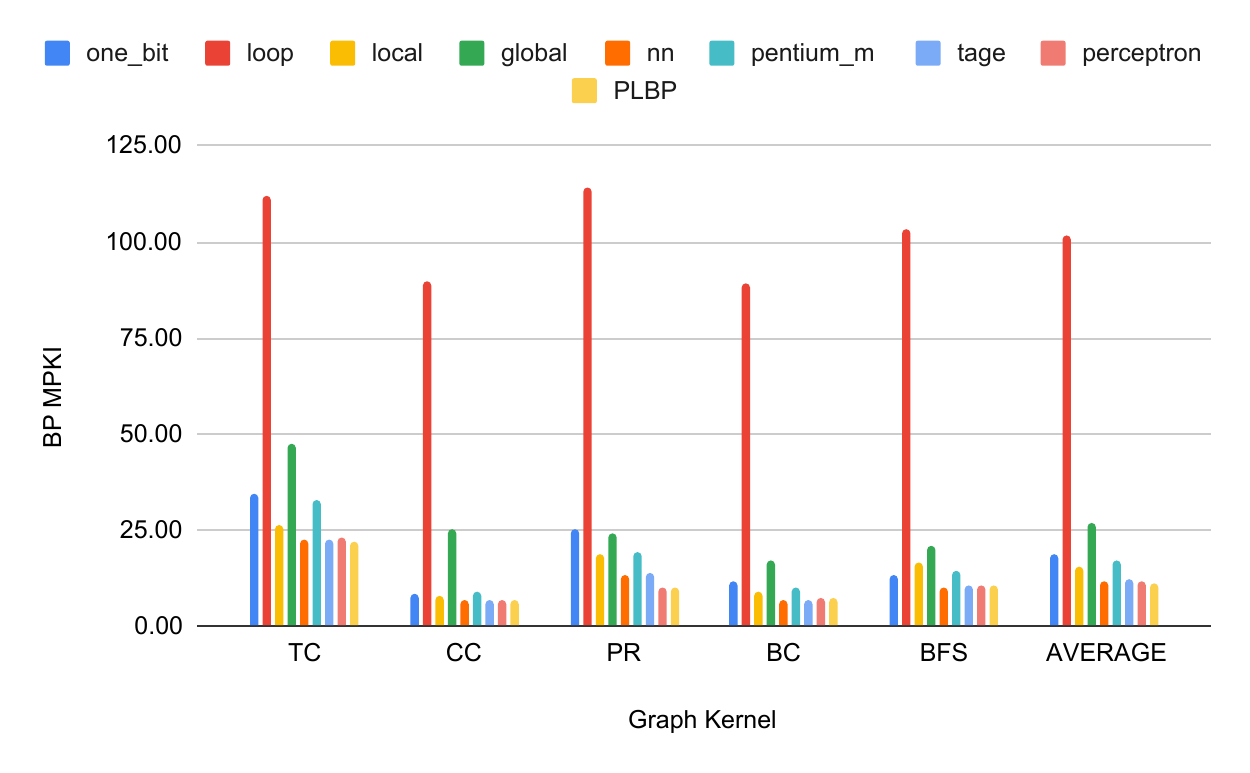}
	\vspace{-2mm}	
	\caption{MPKI of various Branch Predictors for all graph kernels average over datasets amazon, roadCA, webGoogle}
	\label{fig12}
	\vspace{1mm}
\end{figure}

We observe the performance of all the branch predictors discussed in Section I for GAP Benchmark Suite. Fig \ref{fig11} and Fig \ref{fig12} shows the IPC and BP MPKI of various branch predictors. Out of all the predictors used, we find that the loop branch predictor (128 entries, 6 bit tag, 2 ways) shows the minimum performance. Loop predictor detects whether a conditional statement has the loop behavior or not. The loop iteration count in graph applications is dynamic that makes the loop predictor unfit to keep track of dynamic iteration count of a loop statement. Due to this, loop predictor is unable to show much performance improvement. We observe that PLBP branch predictor gives the maximum performance for all the graph kernels as shown in Fig \ref{fig11} and Fig \ref{fig12}, but still less than the perfect branch predictor (BP MPKI for Perfect BP is 0). So, some further optimizations in PLBP can help in improving branch predictor performance for graph applications. PLBP works by learning a set of linear functions for each branch that together comprise a piecewise linear surface. This surface separates the predicted taken branches from the predicted not taken branches \cite{b21} allowing the predictor to predict the linearly inseparable branches. 

\begin{figure}[h]
	\vspace{-2mm}
	\centering
	\includegraphics[width=0.4\textwidth, height=4cm, keepaspectratio, frame]{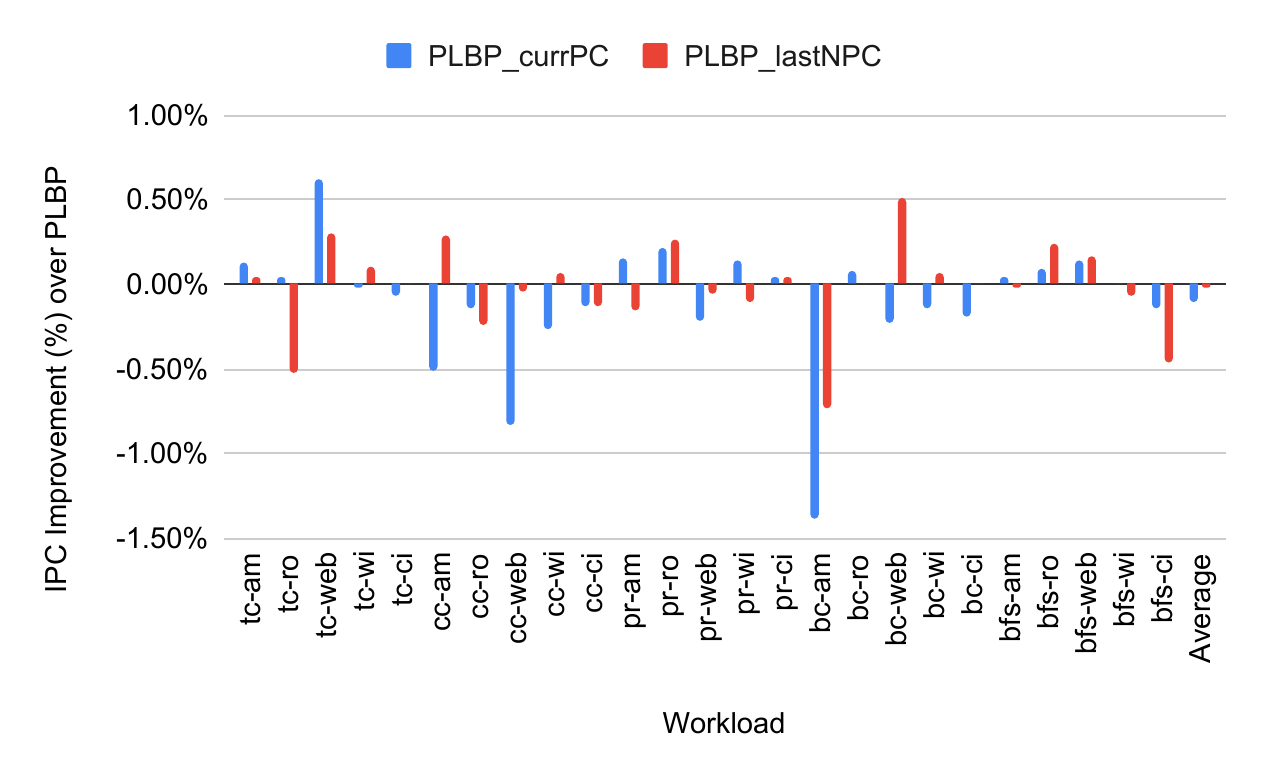}
	\vspace{-2mm}	
	\caption{Performance Improvement of IPC for all graph kernels average over datasets amazon, roadCA, webGoogle, wikitalk, and citepatents using PLBP\_currPC and PLBP\_lastNPC}
	\label{fig13}
	\vspace{1mm}
\end{figure}

\begin{figure}[h]
	\vspace{-2mm}
	\centering
	\includegraphics[width=0.4\textwidth, height=4cm, keepaspectratio, frame]{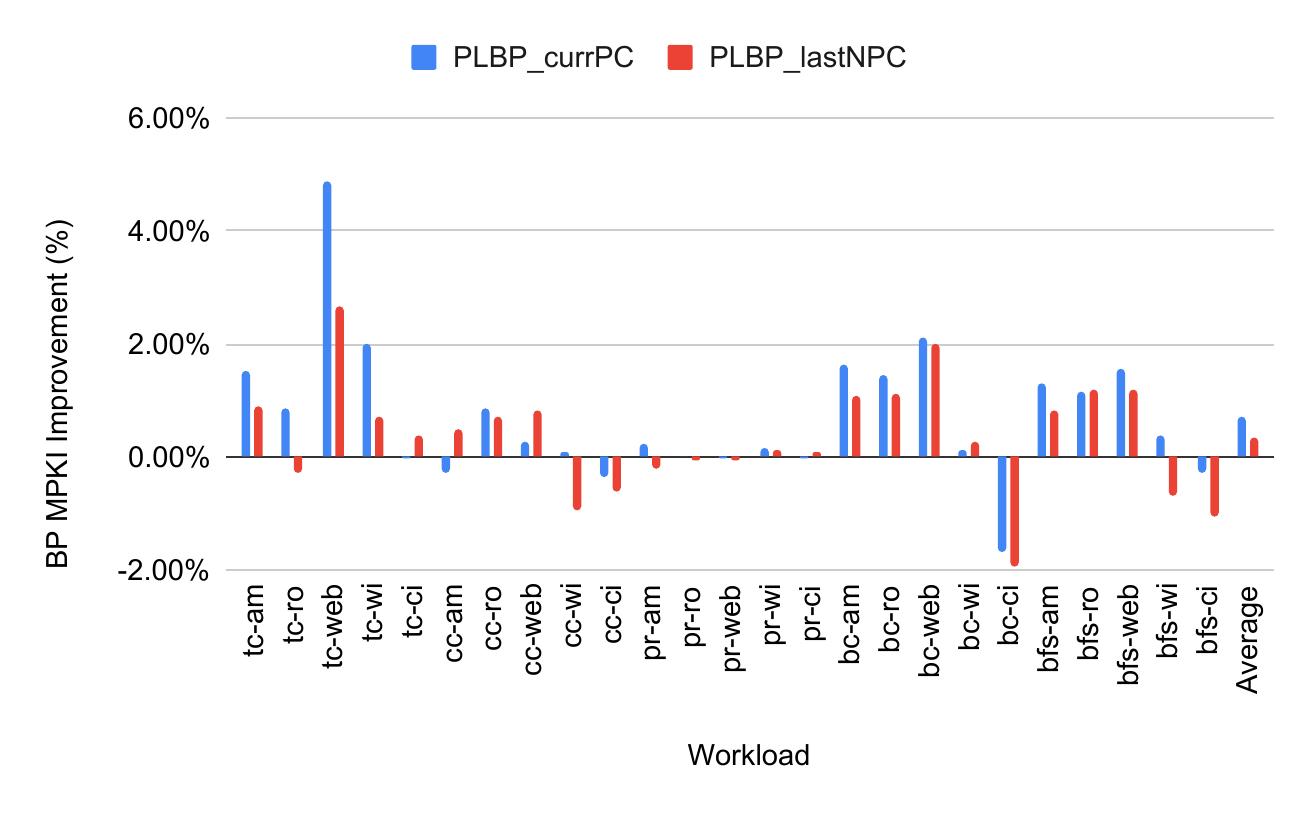}
	\vspace{-2mm}	
	\caption{Performance Improvement of Branch Predictor MPKI for all graph kernels average over datasets amazon, roadCA, webGoogle, wikitalk, and citepatents using PLBP\_currPC and PLBP\_lastNPC}
	\label{fig14}
	\vspace{1mm}
\end{figure}

We further optimize PLBP to uniquely correlate the branch history of a particular branch in the weight table of PLBP by using multiple hash maps and multiple features such as current Program Counter (PC), last N branch PCs, last N PCs, current branch target address, last N branch target address (N=4, set empirically), and adaptive threshold. Out of these, we find the features Branch PC, last 4 Program Counters (lastNPC) and sophisticated hashing mechanism, four\_hybrid12 as better for our optimization. 
We optimize PLBP by using folded XOR operation
on Branch Program Counter (PC), last 4 Program Counters (lastNPC) and four hybrid12 on folded XOR value four\_hybrid12 uses four hash functions namely Wang4shift, Wang3shift, jenkins, and hash7shift \cite{b24} \cite{b25}. We use two optimized versions of PLBP: PLBP\_currPC and PLBP\_lastNPC. PLBP\_currPC uses only current branch PC for folded XOR and  hashing, whereas PLBP\_lastNPC (N=4) uses last 4 PCs including current branch PC for folded XOR and hashing. This hashed value is further used for indexing into the weight table used by PLBP. The IPC and MPKI improvement are for this modified PLBP over baseline PLBP are shown in Fig \ref{fig13} and Fig \ref{fig14}. We observe branch predictor MPKI improvement by 0.71\% and 0.35\% on an average using PLBP\_currPC and PLBP\_lastNPC  with average IPC degradation not more than 0.10\% and 0.2\% respectively.

\subsection{Case study of our work using amazon dataset}
\begin{figure}[h!]
	\vspace{-2mm}
	\centering
	\includegraphics[width=0.4\textwidth, height=4cm, keepaspectratio, frame]{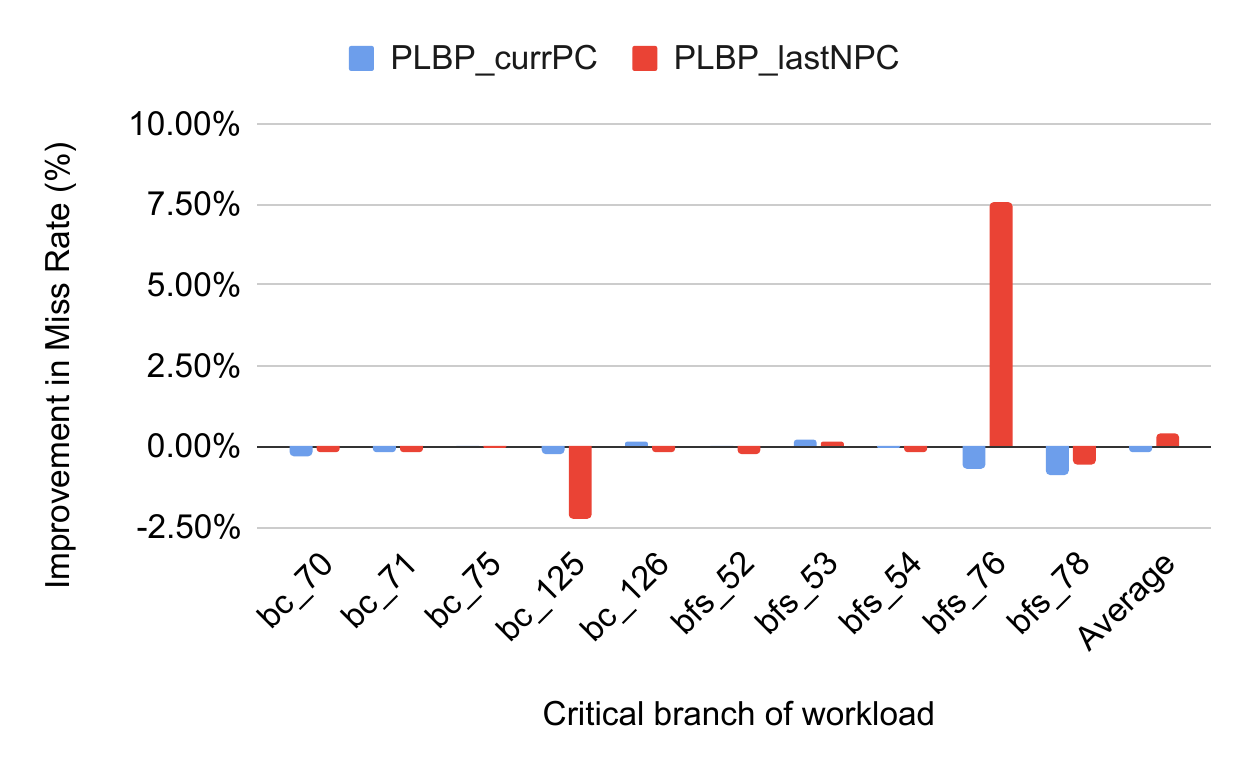}
	\vspace{-2mm}	
	\caption{Miss Rate Improvement (\%) of critical branches showing less impact for PLBP\_currPC and PLBP\_lastNPC for amazon dataset over default PLBP as baseline}
	\label{fig19}
	\vspace{1mm}
\end{figure}

\begin{figure}[h!]
	\vspace{-2mm}
	\centering
	\includegraphics[width=0.4\textwidth, height=4cm, keepaspectratio, frame]{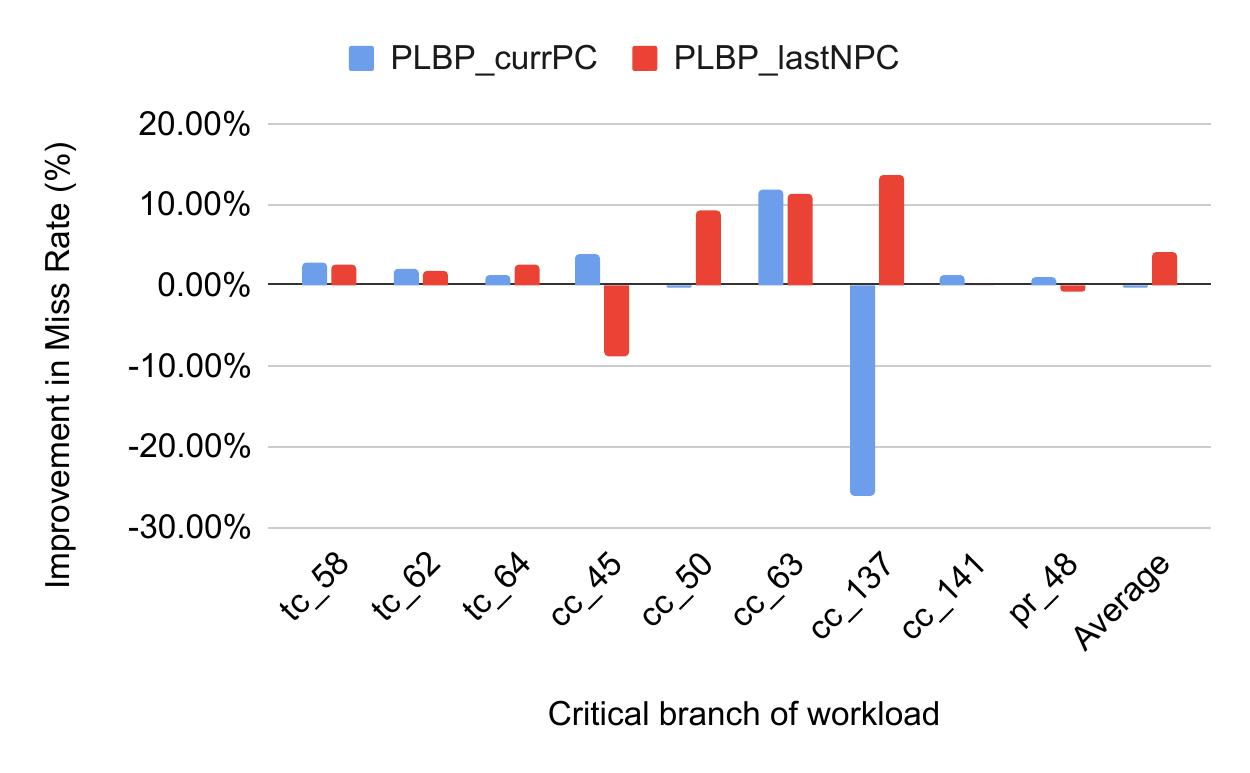}
	\vspace{-2mm}	
	\caption{Miss Rate Improvement (\%) of critical branches showing more impact for PLBP\_currPC and PLBP\_lastNPC for amazon dataset over default PLBP as baseline}
	\label{fig20}
	\vspace{1mm}
\end{figure}

We observe the impact of PLBP\_currPC and PLBP\_lastNPC for individual critical branches over baseline PLBP as shown in Fig \ref{fig19} and Fig \ref{fig20}. Fig \ref{fig19} shows the critical branches that have less impact of PLBP\_currPC and PLBP\_lastNPC over PLBP baseline on miss rate, showing an average improvement of 0.40\% in individual branch miss rate using PLBP\_lastNPC and an average degradation of not more than 0.18\% for PLBP\_currPC. Fig \ref{fig20} shows the critical branches that have more impact of PLBP\_currPC and PLBP\_lastNPC over PLBP baseline on miss rate, showing an average improvement of 4.03\% in individual branch miss rate using PLBP\_lastNPC and an average degradation of not more than 0.40\% for PLBP\_currPC. 

\subsection{Impact of graph reordering on Branch Predictor}
\begin{figure}[h!]
	\vspace{-2mm}
	\centering
	\includegraphics[width=0.40\textwidth, height=4cm, keepaspectratio, frame]{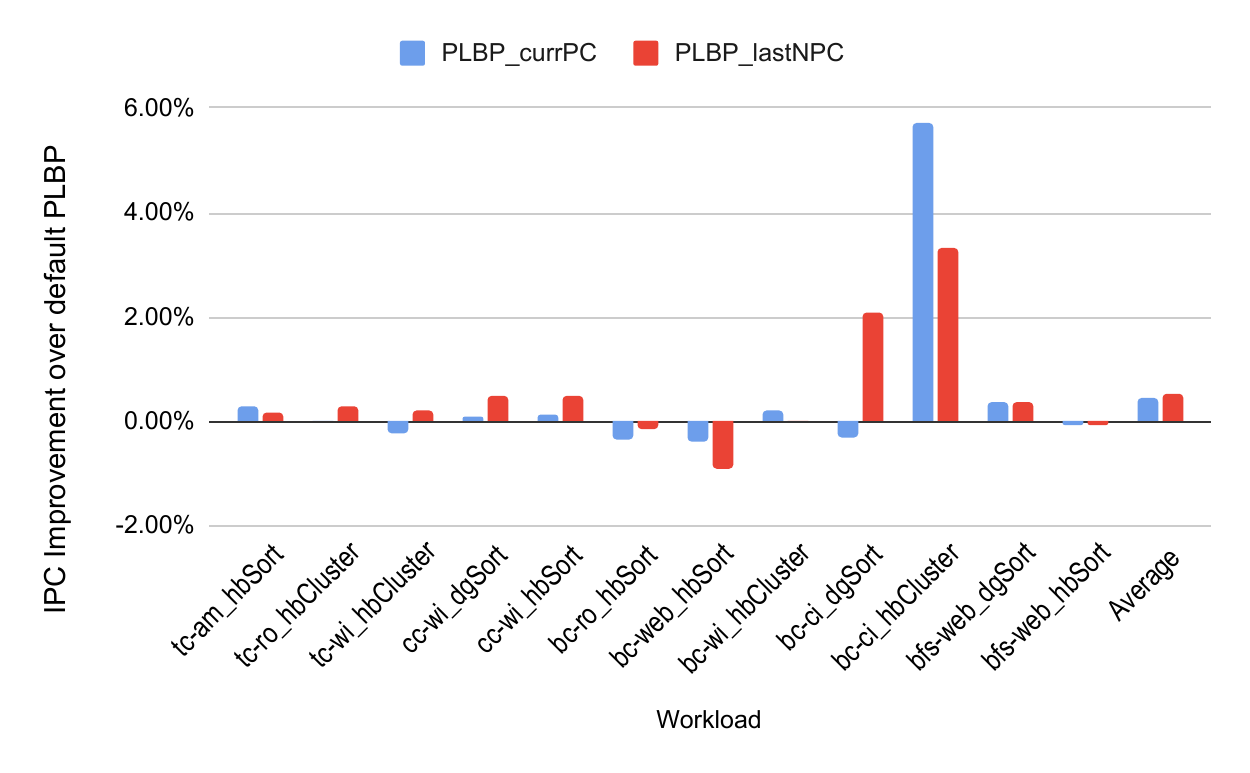}
	\vspace{-2mm}	
	\caption{IPC Improvement (\%) of reordered datasets for PLBP\_currPC and PLBP\_lastNPC over default PLBP as baseline}
	\label{fig21}
	\vspace{1mm}
\end{figure}

\begin{figure}[h!]
	\vspace{-2mm}
	\centering
	\includegraphics[width=0.40\textwidth, height=4cm, keepaspectratio,  frame]{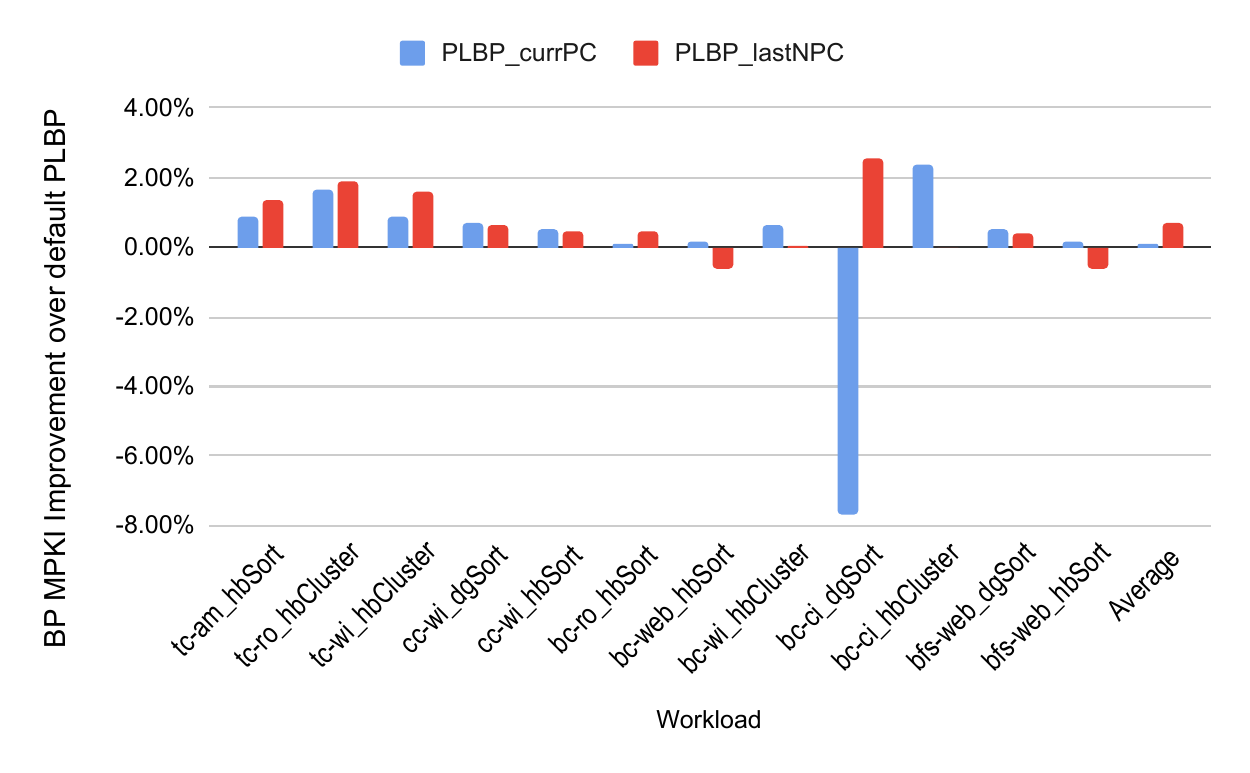}
	\vspace{-2mm}	
	\caption{Branch Predictor MPKI Improvement (\%) of reordered datasets for PLBP\_currPC and PLBP\_lastNPC over default PLBP as baseline}
	\label{fig22}
	\vspace{1mm}
\end{figure}

It is widely observed that graph reordering can significantly improve the locality of data accesses, by grouping the high degree vertices together. The reordering algorithms succeed in bringing the neighbors close to each other, which further increase the reuse of neighbors’ data \cite{b26}. So far, the scope of graph reordering is limited to cache, but we explore the possible benefits of such reordering on branch prediction. We study the impact of graph reordering techniques \cite{b23} such as degree sort, hub sort, and hub clustering on input graphs listed in Table \ref{fig16}. We observe branch predictor IPC improvement of 0.46\% and 0.53\% on an average using PLBP\_currPC and PLBP\_lastNPC  with average MPKI improvement of 0.08\% and 0.68\%  respectively as shown in Fig \ref{fig21} and Fig \ref{fig22}. We find that betweenness centrality algorithm when applied on the largest dataset cite-patents, listed in Table~\ref{fig16} shows maximum improvement in performance due to graph reordering. We observe that the actual improvement in branch predictor performance depends on various factors such as, the specific graph structure (no. of vertices, edges, and degree), the nature of algorithm and its implementation, and the design of branch predictor itself. So, the algorithms exhibiting predictable branch behavior with sorted datasets show reduction in branch predictor MPKI.  

\section{Conclusion}
This work explores multiple branch predictors for graph processing applications. By collecting the branch-wise stats for all the branches, we find that there are very few branches having high misprediction rate as shown in Fig \ref{fig10}. The common critical branches across all the graph workloads are the ones that depend on in-neighbours and out-neighbours of a vertex. Out of various branch predictors, we find that PLBP performs comparatively well on GAPBS workloads. We optimize PLBP by using sophisticated hashing mechanism to reduce the impact of aliasing. We further study the behavior of graph algorithms on sorted datasets and observe that algorithms which have more predictable branch behavior using sorted datasets reduce MPKI and improve overall IPC.

\end{document}